\documentclass[a4paper,11pt]{article}
\usepackage{pos}
\usepackage{multicol}
\usepackage{enumitem}

\newcommand{\beq}{\begin{equation}}
	\newcommand{\eeq}{\end{equation}}

\newenvironment{Figure}
{\par\medskip\noindent\minipage{\linewidth}}
{\endminipage\par\medskip}

\title{Explaining the Cabibbo Angle Anomaly}

\author*{Claudio Andrea Manzari}

\affiliation[]{Physik-Institut, Universit\"at Z\"urich, Winterthurerstrasse 190, CH-8057 Z\"urich, Switzerland}
\affiliation[]{Paul Scherrer Institut, CH-5232 Villigen PSI, Switzerland}

\emailAdd{claudioandrea.manzari@physik.uzh.ch}

\abstract{The Cabibbo-Cobayashi-Maskawa (CKM) matrix parametrizes the misalignement between the up- and down-quark mass basis in the Standard Model (SM). The observation of  first row CKM unitarity violation has recently emerged as a new anomaly of the SM, known as the "Cabibbo Angle Anomaly" (CAA). With current measurements, comparing the elements $V_{ud}$ and $V_{us}$  extracted from beta and kaon decays respectively, the tension with the SM prediction amounts to $\sim $$3\,\sigma$. Recently, it has been pointed out that this anomaly can also be seen as a discrepancy in the determination of the Fermi constant from muon decay vs $\beta$ and K decays, once CKM unitarity is assumed. In fact, possible explanations in terms on New Physics fall under two broad classes: contributions to $\beta$ decay and/or to $\mu$ decay. In this proceedings, we discuss these solutions in terms of gauge invariant dimension 6 operators in SMEFT and simplified extensions of the Standard Model. The latter could introduce correlations with other anomalies in the SM, pointing to new and interesting directions for model building.   
}
\FullConference{%
  *** The European Physical Society Conference on High Energy Physics (EPS-HEP2021), ***\\
  *** 26-30 July 2021 ***\\
  *** Online conference, jointly organized by Universität Hamburg and the research center DESY ***
}

\begin{document}
\maketitle

\section{Introduction}

\indent The observed deficit in first row CKM unitarity, known as the Cabibbo Angle Anomaly (CAA)~\cite{Belfatto:2019swo,Grossman:2019bzp,Coutinho:2019aiy,Manzari:2020eum,Crivellin:2020lzu,Fischer:2021sqw}, is among the most intriguing deviation from the Standard Model (SM) predictions. With the results in Refs.~\cite{Hardy:2020qwl,Marciano:2005ec,Seng:2018yzq,Seng:2018qru,Czarnecki:2019mwq,Seng:2020wjq,Hayen:2020cxh,Shiells:2020fqp,Miller:2008my,Miller:2009cg,Gorchtein:2018fxl} and the compilation of Ref.~\cite{Zyla:2020zbs}, the tension amounts to $\sim$$3\sigma$,
\begin{align}
	|V_{ud}|^2+|V_{us}|^2+|V_{ub}|^2 = 0.9985(5)\,.
	\label{eqCAA}
\end{align}
Here, $V_{ud}$ is extracted from super-allowed $\beta$ decays ($V_{ud}^{\beta}$), $V_{us}$ from $K\to\mu\nu/\pi\to\mu\nu$ and semi-leptonic Kaon decays ($K_{\ell 3}$)  and $V_{ub}$, from B meson decays. 
For this anomaly, $V_{ub}$ is negligible and $|V_{ud}|^2/|V_{us}|^2\approx 20$, such that the sensitivity to NP in the determination of $V_{ud}$ is enhanced with respect to $V_{us}$. In addition, a violation of first column CKM unitarity has also been observed, where again this element dominates but $V_{us}$ is replaced by $V_{cd}$, further strenghtening the idea of NP related to $V_{ud}$. Hence, the presence of New Physics (NP) in the extraction of $V_{ud}$ is preferred to solve the CAA.\\
There are two possible and not necessarly distinct solutions: NP contributions in beta decay and/or in muon decay~\cite{Crivellin:2021njn}. The latter is possible since the Fermi constant, $G_F$, is extracted from muon decay and the product $G_F * V_{ud}$ is measured in beta decays. Within an effective theory approach, these two solutions can be realized by means of operators falling into four classes, as shown in Fig.~\ref{fig:operators}:
\begin{multicols}{2}
	\begin{Figure}
		\includegraphics[width=1.\linewidth]{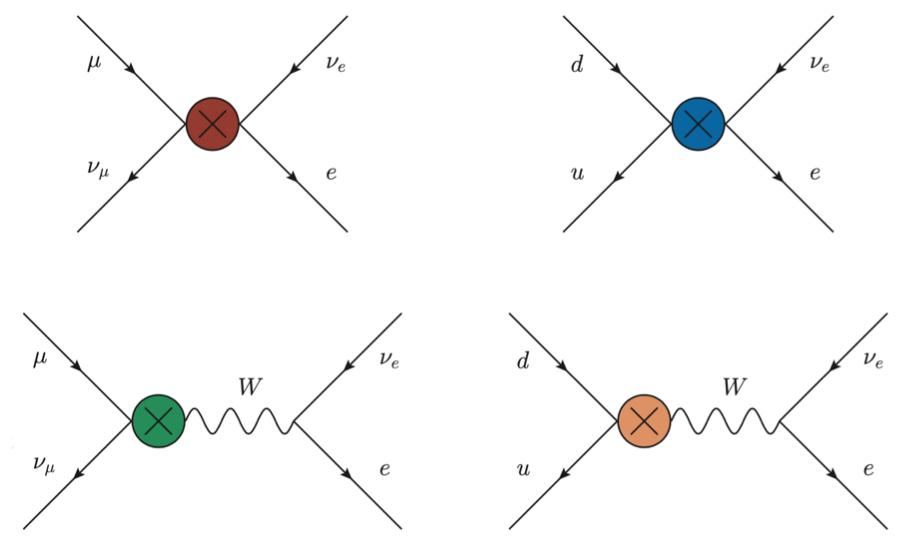}
		\captionof{figure}{Possible classes of solutions to the CAA within an effective theory approach.}
		\label{fig:operators}
	\end{Figure}
	\begin{enumerate}
		\vspace{-1mm}
		\item four-fermion operators in $\mu\to e\nu\nu$,
		\vspace{-1mm}
		\item four-fermion operators in $u\to d e\nu$,
		\vspace{-1mm}
		\item modified $W$--$u$--$d$ couplings,
		\vspace{-1mm}
		\item modified $W$--$\ell$--$\nu$ couplings,
		\vspace{-1mm}
	\end{enumerate}
\end{multicols}
In these proceedings, I discuss the NP effect needed to solve the CAA, first in terms of SMEFT operators at the dimension 6 level and then of simplified SM extensions. Interestingly, the latters introduce correlations with other flavour anomalies, such as  $\tau\to\mu\nu\nu$ and semileptonic B decays. This allows for simultaneous explanations of several tensions and the construction of more appealing SM extensions, highlighting the need for a complete analysis of the CAA.
\\

\section{SMEFT analysis}\label{sec:SMEFT}

As discussed before, to solve the CAA, NP can affect muon and/or beta decay (an additional effect in semi-leptonic $K$ decays would alleviate the tension even further).
\noindent An extensive analysis of all the gauge-invariant dimension-$6$ operators affecting $\beta$ and $\mu$ decay is performed in Ref.~\cite{Crivellin:2021njn}. Here we report the main results, using the conventions of Ref.~\cite{Grzadkowski:2010es}.

\subsection{Four-fermion operators in $\boldsymbol{\mu\to e\nu\nu}$}

Taking into account the constraints from the Michel parameter, muonium-antimuonium oscillations, lepton radiative decays and 3-body lepton flavour violating decays, the only viable way to modify the extraction of $G_F$ proceeds via the SM operator  $Q_{\ell\ell}^{2112} = \big(\bar{\ell}_2\gamma^{\mu}\ell_1\big)\big(\bar{\ell}_1\gamma_{\mu}\ell_2\big)$ (four-quark operators can only contribute via loop effects and are anyway discarded). Because of hermiticity,  its Wilson coefficient is real and can interfere constructively or destructively with the SM in muon decay~\cite{Michel:1949qe,Michel:1954eua,Scheck:1977yg,Fetscher:1986uj,Bayes:2011zza}. In order to bring the tension in Eq.~\ref{eqCAA} at $1\sigma$ we need $C_{\ell \ell }^{2112}\approx -(8\; \rm TeV)^{-2}$. 
This Wilson coefficient is constrained by LEP searches for $e^+e^-\to \mu^+\mu^-$~\cite{Schael:2013ita}. The bounds are a order of magnitude weaker than the value preferred by the CAA, but within reach of future $e^+e^-$ colliders. However, the impact on the electroweak (EW) precision observables has to be taken into account to properly asses the impact of this solution.

\subsection{Four-fermion operators in $\boldsymbol{d\to u e\nu}$}

Here, to have a large enough effect, we need interference with the SM. Taking into account the stringent constraints from $\pi\to\mu\nu/\pi\to e\nu$~\cite{PiENu:2015seu,Zyla:2020zbs}, the scalar operators are ruled out and we are left with $Q_{\ell q}^{\left( 3 \right)1111} = \big(\bar{\ell}_1\gamma^{\mu}\ell_1\big)\big(\bar{q}_1\gamma_{\mu}q_1\big)$.\\
 The CAA prefers $C_{\ell q}^{\left( 3 \right)1111}\approx (10\; \rm TeV)^{-2}$.
Via $SU(2)_L$ invariance, this operator generates effects in neutral-current interactions and at constructive interference with the $\bar uu (\bar dd)\to e^+e^-$ amplitudes at high energies. The limits extracted from non-resonant di-lepton searches by ATLAS~\cite{Aad:2020otl} are still non-onstraining but start to be important in the case of a triplet operator. Interesting correlations with the latest CMS observations in non-resonant di-lepton searches have been discussed in Ref.~\cite{Crivellin:2021rbf}.

\subsection{Modified $\boldsymbol{W}$--$\boldsymbol{u}$--$\boldsymbol{d}$ couplings}

There are only two operators modifying the $W$ couplings to quarks	$Q_{\phi q}^{\left( 3 \right)ij} = {\phi ^\dag }i\overset{\leftrightarrow}{D}^I_\mu\phi {{\bar q}_i}{\gamma ^\mu }{\tau ^I}{q_j}$ and $Q_{\phi ud}^{ij} = {\tilde \phi ^\dag }iD_\mu \phi {{\bar u}_i}{\gamma ^\mu }{d_j}$.
First of all, $Q_{\phi ud}^{11(12)}$ generates right-handed $W$--quark couplings~\cite{Buras:2010pz}, which can solve the CAA, accounting also for the difference between $K_{\ell2}$ and $K_{\ell3}$ decays~\cite{Grossman:2019bzp}. A solution with $Q_{\phi q}^{\left( 3 \right)11}$ is possible, modifying the left-handed $W$--quark couplings and data prefer $C_{\phi q}^{\left( 3 \right)11} \approx -(9\; \rm TeV)^{-2}$.
Due to $SU(2)_L$ invariance, constraints from $D^0-\bar{D}^0$ and $K^0-\bar{K}^0$ mixing and from $Z$ decays have to be taken into account. However, the former bounds can be avoided by a $U(2)$ flavor symmetry and the latter by simultaneous contributions to $Q_{\phi q}^{\left( 1 \right)ij}$. For a detailed analysis see Ref.~\cite{Belfatto:2021jhf}.

\subsection{Modified $\boldsymbol{W}$--$\boldsymbol{\ell}$--$\boldsymbol{\nu}$ couplings}

Only the operator $	Q_{\phi \ell }^{(3)ij}={\phi ^\dag }i\overset{\leftrightarrow}{D}^I_\mu\phi {{\bar\ell}_i}{\gamma ^\mu }{\tau ^I}{\ell_j}$ generates modified $W$--$\ell$--$\nu$ couplings at tree level.
The off-diagonal Wilson coefficients are neglected to avoid the stringent bounds from charged lepton flavor violation and since they generate negligible effects. Modified $W$ couplings to electrons affect muon and beta decay in the same way and leave the CAA unaffected. On the other hand, $C_{\phi \ell }^{(3)22}$ only enters in muon decay and provides a viable solution to the anomaly, which prefers $C_{\phi \ell }^{(3)22}> 0$. A non-zero $C_{\phi \ell }^{(3)11}<0$, with $|C_{\phi \ell }^{(3)22}|<|C_{\phi \ell }^{(3)11}|$, is also required by lepton flavor universality tests such as $\pi,\, K$ and $\tau$ decays~\cite{Coutinho:2019aiy,Crivellin:2020lzu}. Also in this case, a global fit including the impact on the EW precision observables is required.

\section{Simplified Models}

An exhaustive analysis of simplified models which can generate the operators discussed in Sec.~\ref{sec:SMEFT} is beyond the scope of this proceedings. Here, we report the cases studied in the literature with the focus on emerging correlations with other anomalies. The possible SM extensions can be sum up by their effects as follows: 

\setlength\multicolsep{\topsep}
\begin{multicols}{2}
\centering 
\textbf{NP in $\pmb \mu$ decay}
	\begin{enumerate}
	\item Singly Charged Scalar Singlet
	\vspace{-1mm}
	\item Vector Boson Singlet
	\vspace{-1mm}
	\item Vector Boson Triplet
	\vspace{-1mm}
	\item Vector-like Leptons
\end{enumerate}
\columnbreak
\textbf{NP in $\pmb \beta$ decay}
	\begin{enumerate}
	\item Vector Boson Triplet
	\vspace{-1mm}
	\item Vector-like Quarks
	\vspace{-1mm}
	\item Vector-like Leptons
	\item Leptoquarks 
	\end{enumerate}
\end{multicols}
\noindent A neutral vector boson $SU(2)_L$ singlet can contribute to the muon decay amplitude. Taking into account the constrains from charged lepton flavour violating decays, EW precision observables and LEP bounds, only LFV couplings can solve the CAA~\cite{Buras:2021btx}.\\
On the other hand, a vector boson triplet allows a very simple solution of the CAA via the $W^{\prime}$ contribution to $\beta$ and $\mu$ decays simultaneously. In addition, the neutral component of the triplet could introduce interesting correlations with $b\to s\ell\ell$ observables. In fact, Ref.~\cite{Capdevila:2020rrl} shows that this model can provide a solution to the CAA and strongly improve the agreement with data in semileptonic B decays.\\
Solutions to the anomaly with vector-like quarks (VLQ) and vector-like leptons (VLL) are discussed in Ref.~\cite{Belfatto:2019swo,Grossman:2019bzp,Kirk:2020wdk,Crivellin:2020ebi,Manzari:2021prf,Belfatto:2021jhf}. They proceed via a modification of the gauge boson couplings with quarks and leptons respectively, therefore a global fit to a large set of observables has to be performed. With VLQs two solutions are possible~\cite{Kirk:2020wdk,Fischer:2021sqw}: one involving two singlets coupling to up and down quarks respectively and one involving a doublet coupling to them simultaneously. Interestingly, Ref.~\cite{Crivellin:2020ebi} finds a VLL model, made of a singlet coupling with electrons and a triplet coupling with muons, which solve the CAA and significantly improves the global fit to data.\\
Leptoquark solutions have been discussed in Refs.~\cite{Crivellin:2021egp,Crivellin:2021bkd}, where it is shown that the scalar triplet can address the anomaly once an additional mechanism to compensate the effect in $D^0-\bar{D^0}$ is introduced.  
Eventually, there is the Singly Charged Scalar Singlet, which is a $SU(2)_L\times SU(3)_C$ singlet with hypercharge $+1$. Because of its quantum numbers, it cannot couple to quarks and naturally introduces lepton flavour violation, contributing to muon decay amplitude and  providing a solution to the CAA~\cite{Crivellin:2020klg}. In Ref.~\cite{Crivellin:2020oup} it is used in conjunction with a neutral vector boson singlet and VLQs to explain also $b\to s\ell\ell$, $Z\to \bar{b}b$ and $\tau\to\mu\nu\nu$ data, while in Ref.~\cite{Marzocca:2021azj} together with a scalar leptoquark to explain also $(g-2)_{\mu}$ and charged-current B decays.

\section{Conclusion}

The observed first row CKM unitarity violation, at the $3\, \sigma$ level, has recently been the object of a detailed study. Interpreting it as a new hint of LFU violation allows for simple solutions of the tension and introduces intriguing correlations with other aomalies in the flavour sector. An analysis in terms of effective operators identifies 4 classes of possible solutions: four-fermion operators affecting $\mu\to e\nu\nu$, four-fermion operators affecting $d\to ue\nu$, modified $W$--$u$--$d$ couplings and modified $W$--$\ell$--$\nu$ couplings.  
Taking into account the constraints from all relevant observables, only five gauge invariant dim-6 operators, in SM effective field theory (SMEFT), can account for the CAA anomaly: $Q_{\ell\ell}^{2112},\, Q_{\ell q}^{\left( 3 \right)1111},\, Q_{\phi ud}^{11(12)},\, Q_{\phi q}^{\left( 3 \right)11}$ and $Q_{\phi \ell }^{(3)22}$.\\
Then, performing a systematic study of simplified models which, one finds that a vector boson singlet, a vector boson triplet, a singly charged scalar, vector-like leptons, vector-like quarks and leptoquarks can generate these operators. The vector boson singlet and the leptoquarks can only allevatiate the CAA while all the other simplified models can completely solve it. In addition, they can give rise to interesting correlations with other observables such as $b\to s\ell\ell,\, Z\to\bar{b}b$ and $\tau\to\mu\nu\nu$. This result suggests a common explanatory framework for several anomalies and opens up novel and interesting avenues for model building. At the same time, the need for developments in the extraction of $V_{us}$ and $V_{ud}$ and in the analysis of the CAA, on the theoretical and experimental side, is highlighted.

\bibliographystyle{JHEP}
\bibliography{GF}

\end{document}